\documentclass[twocolumn,showpacs,aps,prd]{revtex4}

\usepackage{epsfig}
\usepackage{graphics}
\usepackage{graphicx}
\usepackage{epic}
\usepackage{floatflt}
\usepackage{dcolumn}
\usepackage{amsmath}

\RequirePackage{xspace}

\usepackage{relsize}
\def\babar{\mbox{\slshape B\kern-0.1em{\smaller A}\kern-0.1em
    B\kern-0.1em{\smaller A\kern-0.2em R}}}

\def\Kbar  {\kern 0.2em\overline{\kern -0.2em K}{}\xspace}

\def\KS    {\ensuremath{K_{\scriptscriptstyle S}}\xspace} 
 
\def\Kstarz  {\ensuremath{K^{*0}}\xspace}
\def\Kstarzb  {\ensuremath{\Kbar^{*0}}\xspace}

\def\Kzb   {\ensuremath{\Kbar^0}\xspace}
\def\KzKzb {\ensuremath{K^0 \kern -0.16em \Kzb}\xspace}
\def\Dz    {\ensuremath{D^0}\xspace}

\def\Dbar  {\kern 0.2em\overline{\kern -0.2em D}{}\xspace}

\def\Dzb   {\ensuremath{\Dbar^0}\xspace}
\def\DzDzb {\ensuremath{D^0 {\kern -0.16em \Dzb}}\xspace}

\def\Bz    {\ensuremath{B^0}\xspace}
\def\B     {\ensuremath{B}\xspace}
\def\Bbar  {\kern 0.18em\overline{\kern -0.18em B}{}\xspace}

\def\Bzb   {\ensuremath{\Bbar^0}\xspace}
\def\Bu    {\ensuremath{B^+}\xspace}
\def\Bub   {\ensuremath{B^-}\xspace}

\def\BzBzb {\ensuremath{B^0 {\kern -0.16em \Bzb}}\xspace}
\def\BpBm    {\ensuremath{\Bu {\kern -0.16em \Bub}}\xspace}

\mathchardef\Upsilon="7107
\def\Y#1S{\ensuremath{\Upsilon{(#1S)}}\xspace}

\def\FourS {\Y4S}

\mathchardef\Deltares="7101
\mathchardef\Xi="7104
\mathchardef\Lambda="7103
\mathchardef\Sigma="7106
\mathchardef\Omega="710A
\def\Deltabar   {\kern 0.25em\overline{\kern -0.25em \Deltares}{}\xspace}
\def\Lbar {\kern 0.2em\overline{\kern -0.2em\Lambda\kern 0.05em}\kern-0.05em{}\xspace}
\def\Sigbar{\kern 0.2em\overline{\kern -0.2em \Sigma}{}\xspace}
\def\Xibar{\kern 0.2em\overline{\kern -0.2em \Xi}{}\xspace}
\def\Obar{\kern 0.2em\overline{\kern -0.2em \Omega}{}\xspace}
\def\Nbar{\kern 0.2em\overline{\kern -0.2em N}{}\xspace}
\def\Xb{\kern 0.2em\overline{\kern -0.2em X}{}}

\def\mes        {\mbox{$m_{\rm ES}$}\xspace}

\newcommand{\tev}{\ensuremath{\mathrm{\,Te\kern -0.1em V}}\xspace}
\newcommand{\gev}{\ensuremath{\mathrm{\,Ge\kern -0.1em V}}\xspace}
\newcommand{\mev}{\ensuremath{\mathrm{\,Me\kern -0.1em V}}\xspace}
\newcommand{\kev}{\ensuremath{\mathrm{\,ke\kern -0.1em V}}\xspace}
\newcommand{\ev}{\ensuremath{\mathrm{\,e\kern -0.1em V}}\xspace}
\newcommand{\gevc}{\ensuremath{{\mathrm{\,Ge\kern -0.1em V\!/}c}}\xspace}
\newcommand{\mevc}{\ensuremath{{\mathrm{\,Me\kern -0.1em V\!/}c}}\xspace}
\newcommand{\gevcc}{\ensuremath{{\mathrm{\,Ge\kern -0.1em V\!/}c^2}}\xspace}
\newcommand{\mevcc}{\ensuremath{{\mathrm{\,Me\kern -0.1em V\!/}c^2}}\xspace}

\def\mus  {\ensuremath{\rm \,\mus}\xspace}

\def\mus        {\ensuremath{\,\mu{\rm s}}\xspace}    
\def\gsim{{~\raise.15em\hbox{$>$}\kern-.85em
          \lower.35em\hbox{$\sim$}~}\xspace}
\def\lsim{{~\raise.15em\hbox{$<$}\kern-.85em
          \lower.35em\hbox{$\sim$}~}\xspace}

\def\CP                 {\ensuremath{C\!P}\xspace}

\def\to                 {\ensuremath{\rightarrow}\xspace}

\def\pep2{PEP-II}

\def\deltaz{\ensuremath{{\rm \Delta}z}\xspace}
\def\deltat{\ensuremath{{\rm \Delta}t}\xspace}

\xspace

\newcommand{\pr}        [1]  {{Phys.\ Rev.\ {\bf #1}}}

\newcommand{\annp}  [1]  {{Ann.\ Phys.\ {\bf #1}}}

\def\jetset74   {\mbox{\tt Jetset \hspace{-0.5em}7.\hspace{-0.2em}4}}

\def\bea{\begin{eqnarray}}
\def\eea{\end{eqnarray}}

\def\mes{\ensuremath{m_{\rm ES}}}
\def\de{\ensuremath{\Delta E}}
\def\F{\ensuremath{{\cal F}}}

\def\sig{\ensuremath{{\sf Sig}}}

\def\bb{\ensuremath{{\sf B\overline{B}}}}

\def\de {\ensuremath{\Delta E}}

\def\Kstarz {K^{*0}}

\def\bztdzksz {\ensuremath{B^0 \rightarrow {\Dtilde}^{0}K^{*0}}}

\def\bea{\begin{eqnarray}}
\def\eea{\end{eqnarray}}

\def\mes        {\mbox{$m_{\rm ES}$}\xspace}

\newcommand{\fis}{\ensuremath{\mbox{$\mathcal{F}$}}}
\def\de{\ensuremath{\Delta E}}
\def\F{\ensuremath{{\cal F}}}

\def\sig{\ensuremath{{\sf Sig}}}

\def\bb{{{B\bar{B}}}}

\def\Dtilde {\ensuremath{\tilde{D}}\xspace}
\newcommand{\BaBarType}      {PUB}  
\newcommand{\BaBarYear}       {07}
\newcommand{\BaBarNumber}     {072}
\newcommand{\SLACPubNumber} {13229}

\begin{document}

\begin{flushleft}
\babar-\BaBarType-\BaBarYear/\BaBarNumber \\ 
SLAC-PUB-\SLACPubNumber\\ 
\end{flushleft}
\title{Constraints on the CKM angle $\gamma$ in $\Bz\to\Dzb\Kstarz$ and $\Bz\to\Dz\Kstarz$ 
from a Dalitz analysis of $\Dz$ and $\Dzb$ decays to $K_S\pi^+\pi^-$}  

%
\author{B.~Aubert}
\author{M.~Bona}
\author{Y.~Karyotakis}
\author{J.~P.~Lees}
\author{V.~Poireau}
\author{X.~Prudent}
\author{V.~Tisserand}
\author{A.~Zghiche}
\affiliation{Laboratoire de Physique des Particules, IN2P3/CNRS et Universit\'e de Savoie, F-74941 Annecy-Le-Vieux, France }
\author{J.~Garra~Tico}
\author{E.~Grauges}
\affiliation{Universitat de Barcelona, Facultat de Fisica, Departament ECM, E-08028 Barcelona, Spain }
% ...
\collaboration{The \babar\ Collaboration}
\noaffiliation

\date{\today}
\begin{abstract}
\noindent
We present constraints on the angle $\gamma$ of the Unitarity Triangle with 
a Dalitz analysis of neutral $D$ decays to 
$K_S\pi^+\pi^-$ from the processes $\Bz \rightarrow \Dzb\Kstarz$ 
($\Bzb \rightarrow \Dz\Kstarzb$) and $\Bz \rightarrow \Dz\Kstarz$ ($\Bzb \rightarrow \Dzb\Kstarzb$) 
with $\Kstarz \rightarrow 
K^{+}\pi^{-}$ ($\Kstarzb \rightarrow K^{-} \pi^{+}$). 
Using a sample of $371\times 10^6$ $\bb$ pairs collected with the \babar\ 
detector at PEP-II, 
we constrain the angle $\gamma$ as a function of $r_{S}$, the magnitude of the 
average ratio between $b \rightarrow u$ and $b \rightarrow c$ amplitudes. 

\end{abstract}
\maketitle
\section{Introduction}
Various methods have been proposed to
determine the Unitarity Triangle angle $\gamma$~\cite{ref:GLW,ref:ADS,ref:DKDalitz} 
of the Cabibbo-Kobayashi-Maskawa (CKM) quark mixing matrix \cite{ref:ckm} using 
$\ensuremath{B^{-} \to \Dtilde^{(*)0}K^{(*)-}}$ decays, where the symbol $\Dtilde^{(*)0}$ 
indicates either a $D^{(*)0}$ or a $\Dbar^{(*)0}$ meson. 
A $B^-$ can decay into a $\Dtilde^{(*)0}K^{(*)-}$ 
final state via a $b\to c$ or a $b\to u$ mediated process and \CP\ violation 
can be detected when the $D^{(*)0}$ and the $\Dbar^{(*)0}$ decay to the same final state.  
These processes are thus sensitive to $\gamma=\mbox{arg}\{-V^{*}_{ub}V_{ud}/V^{*}_{cb}V_{cd}\}$. 
The present determination of $\gamma$ comes from the combination of
several results obtained with the different methods.
In particular, the Dalitz technique \cite{ref:DKDalitz}, when used to analyze
$\ensuremath{B^{-} \to \Dtilde^{(*)0}K^{(*)-}}$ decays, is very powerful,
resulting in an error on $\gamma$ of about
24$^\circ$ and 13$^\circ$ for the \babar\ and Belle analyses respectively
(\cite{ref:DKDalitzbabar,ref:BelleDalitz}). These results are obtained
from the simultaneous exploitation of the three decays of the charged
$B$ mesons ($\ensuremath{B^{-} \to \Dtilde^{0}K^{-}}$, $\Dtilde^{*0}K^{-}$
and $\Dtilde^{0}K^{*-}$) and, in the case of \babar, from the study of
two final states for the neutral $D$ mesons ($K_S\pi^+\pi^-$ and $K_S K^+ K^-$).

\begin{figure}[h!]
\begin{center}
\epsfig{file=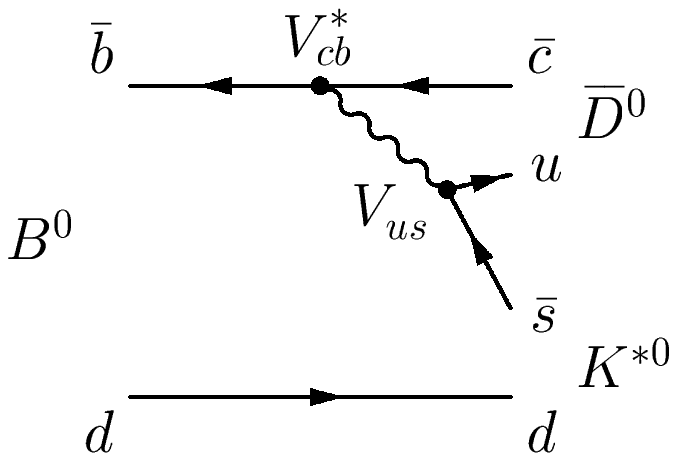,width=0.45\linewidth}
\epsfig{file=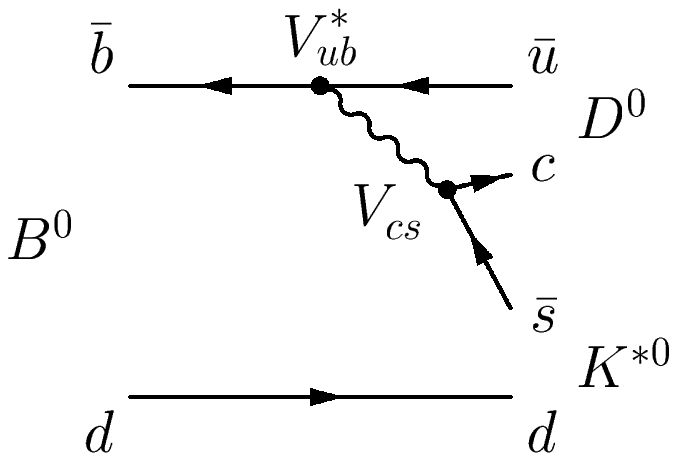,width=0.45\linewidth}\\
\epsfig{file=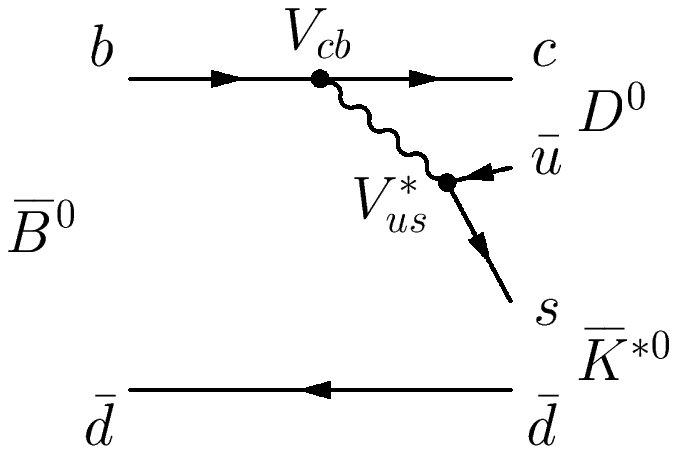,width=0.45\linewidth}
\epsfig{file=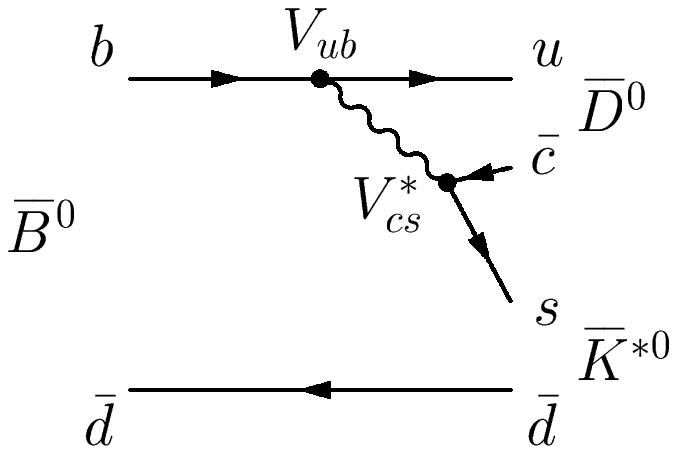,width=0.45\linewidth}
\end{center}
\caption{Feynman diagrams for the decays $B^0 \rightarrow \Dzb \Kstarz$ (up left, 
$\bar b \to \bar c$ transition), $B^0 \rightarrow \Dz \Kstarz$ 
(up right, $\bar b \to \bar u$ transition), $\overline{B}^0 \rightarrow \Dz \Kstarzb$ (down left,
$b \to c$ transition), and $\overline{B}^0 \rightarrow \Dzb \Kstarzb$ (down right, $ b \to u$ transition).  
A $\Kstarz$ is a decay product of a $\Bz$ while a $\Kstarzb$ results from a $\Bzb$ decay.}
\label{fig:feyn}
\end{figure}

In this paper we present the first measurement of the angle $\gamma$ 
using neutral $B$ meson decays.
We reconstruct $\Bz \rightarrow \Dtilde^0 \Kstarz $, with 
$\Kstarz \to K^+ \pi^-$ (charge conjugate processes are assumed throughout the 
paper and $\Kstarz$ refers to $K^{*}(892)^{0}$), where the flavor of the $B$ meson is identified by 
the kaon electric charge. 
Neutral $D$ mesons are reconstructed in the $\KS\pi^{+}\pi^{-}$ decay mode and
are analyzed with the Dalitz technique \cite{ref:DKDalitz}. 
The final states we reconstruct can be reached through $b \to c$ and $b \to u$ processes with the 
diagrams shown in Fig.~\ref{fig:feyn}.  
The correlation within the flavor of the neutral $D$ meson and the charge of the kaon in the 
final state allows for discriminating between events arising from $b\to c$ and $b\to u$ transitions. 
In particular it is useful for the following discussion to stress that $\bar b \to \bar u$ 
($B^0$) transitions lead to $\Dz K^+\pi^-$ final states and $b\to u$ ($\overline{B}^0$) transitions 
lead to $\Dzb K^-\pi^+$ final states.

When analyzing $\Bz \rightarrow \Dtilde^0 \Kstarz $ decays, the 
natural width of the $K^{*0}$ (50 MeV/$c^2$) has to be 
considered. In the $K^{*0}$ mass region, amplitudes for decays to
higher-mass $K\pi$ resonances interfere with the signal decay 
amplitude and with each other.
For this analysis we use effective variables, introduced in 
Ref.~\cite{gronau2002}, obtained by integrating 
over a region of the $B^0\to\Dtilde^0 K^+ \pi^-$ Dalitz plot 
corresponding to the $\Kstarz$ .
For this purpose we introduce the quantities $r_S$, $k$, 
and $\delta_S$ defined as
\begin{eqnarray}
&& r_S^2\equiv\frac{\Gamma(\Bz \to \Dz K^+\pi^-)}{\Gamma(\Bz \to \Dzb K^+\pi^-)}
=\frac{\int dp\ A_{u}^2(p)}{\int dp\ A_{c}^2(p)},\label{eq:rs_square}\\
&& ke^{i\delta_S}\equiv\frac{\int dp\ A_{c}(p)A_{u}(p)e^{i\delta (p)}}{\sqrt{\int dp\ A_{c}^2(p) \ \int dp\ A_{u}^2(p)}}\, ,\label{eq:k}
\end{eqnarray}
where $0\leq k \leq 1$ and $\delta_S\in[0,2\pi]$.
The amplitudes for the $b\to c$ and $b\to u$ transitions, $A_{c}(p)$ and $A_{u}(p)$, 
are real and positive and $\delta(p)$ is the relative strong phase. 
The variable $p$ indicates the position in the $\Dtilde^0 K^+ \pi^-$ Dalitz plot.
In case of a two-body $\B$ decay, $r_S$ and $\delta_S$ become $r_B=|A_{u}|/|A_{c}|$ 
and $\delta_B$ (the strong phase difference between $A_{u}$ and $A_{c}$) and $k=1$.
Because of CKM factors and the fact that both diagrams, for the neutral $B$ 
decays we consider, are color-suppressed,
the average amplitude ratio $r_S$ in \bztdzksz\ is expected to be in the range [0.3, 0.5], 
larger than the analogous ratio 
for charged $B^{\pm}\to \Dtilde^0K^{\pm}$ decays (which is of the order of $10\%$~\cite{Bona:2005vz,ref:onlyforfairness}). 
An earlier measurement 
sets an upper limit $r_S  < 0.4 ~\rm{at}~ 90\%$ probability \cite{ref:sha}.
A phenomenological approach \cite{ref:cavoto} proposed to evaluate $r_B$  in the 
$B^0 \rightarrow \Dtilde^0 K^0$ system gives $r_B=$0.27$\pm$0.18.
\section{Event Reconstruction and Selection}
The analysis presented in this paper uses a data sample of 
$371\times 10^{6}$ $\bb$ pairs collected with the \babar\ detector at 
the \pep2\ storage ring. Approximately 10\% of the collected data
(35~fb$^{-1}$) have a center-of-mass (CM) energy 40~\mev below the
$\FourS$ resonance.
These ``off-resonance'' data are used to study backgrounds from continuum
events, $e^+ e^- \to q \bar{q}$ ($q=u,d,s,$ or $c$).  

The \babar\ detector is described elsewhere~\cite{ref:det}. 
Charged-particle tracking is provided by a five-layer silicon 
vertex tracker (SVT) and a 40-layer drift chamber (DCH).  
In addition to providing precise position information for tracking, the SVT and DCH 
also measure the specific ionization ($dE/dx$), which is used for particle 
identification of low-momentum charged particles. At higher momenta ($p>0.7$~\gevc) 
pions and kaons are identified by Cherenkov radiation detected in a ring-imaging 
device (DIRC).  
The position and energy of photons are 
measured with an electromagnetic calorimeter (EMC) consisting of 6580 
thallium-doped CsI crystals.  
These systems are mounted inside a 1.5T solenoidal super-conducting magnet.  

We reconstruct $\Bz\to\Dtilde^0\Kstarz$ events with $\Kstarz\to K^{+}\pi^-$ and 
$\Dtilde^0\to\KS\pi^+\pi^-$. The event selection, described below, is developed 
from studies of off-resonance data and events simulated with Monte Carlo 
techniques (MC). 
The \KS\ is reconstructed from pairs of oppositely-charged pions with 
invariant mass within 7 \mevcc\ of the nominal \KS\ mass \cite{ref:PDG}, 
corresponding to 2.8 standard deviations of the mass distribution for signal events.
We also require that $\cos \alpha_{\KS}(\Dtilde^0)>0.997$, where $\alpha_{\KS}(\Dtilde^0)$ is the 
angle between the \KS\ 
line of flight (line between the $\Dtilde^0$\ and the \KS\ decay points) and the \KS\ 
momentum (measured from the two pion momenta).
Neutral $D$ candidates are selected by combining \KS\ candidates with two oppositely-charged 
pion candidates and requiring the $\Dtilde^0$ invariant mass to be within 11 \mevcc\ of 
its nominal mass \cite{ref:PDG}, corresponding to 1.8 standard deviations of the mass 
distribution for signal events.  
The \KS\ and the two pions used to reconstruct the $\Dtilde^0$ are constrained to originate 
from a common vertex. 
The charged kaon is required to satisfy kaon identification criteria, which are 
based on Cherenkov angle and $dE/dx$ measurements and are typically 85\% efficient, 
depending on momentum and polar angle.  Misidentification rates are at the 2\% level.  
The tracks used to reconstruct the $\Kstarz$ are constrained to originate 
from a common vertex and their invariant mass is required to 
lie within 48 \mevcc of the nominal $\Kstarz$ mass \cite{ref:PDG}. 
We define $\theta_{Hel}$ as the angle between the direction of flight of 
the charged $K$ in the $\Kstarz$ rest frame with respect to the direction 
of flight of the $K^{*0}$ in the \B\ rest frame.  
The distribution of $\cos\theta_{Hel}$ is expected to be proportional to $\cos^2\theta_{Hel}$ for 
signal events, due to angular momentum conservation, and flat for background events.  
We require $|\cos\theta_{Hel}|>0.3$.  
The cuts on the $K^{*0}$ mass and on $|\cos\theta_{Hel}|$ have been 
optimized maximizing the function $S/\sqrt{S+B}$, where $S$ and $B$ are the 
expected numbers of signal and background events respectively, based on 
MC studies.  
The $B^0$ candidates are reconstructed by combining one $\Dtilde^0$ 
and one $K^{*0}$ candidate, constraining them to originate from a 
common vertex with a probability greater than 0.001. 
The distribution of the cosine of the $B$ polar angle with respect to the 
beam axis in the $e^+e^-$ CM frame, $\cos\theta_{B}$, is expected to be 
proportional to $1-\cos^2\theta_B$.  We require $|\cos\theta_{B}|<0.9$.  

We measure two almost independent kinematic variables: the 
beam-energy substituted mass 
$\mes\equiv\sqrt{(E^{*2}_{0}/2+\vec{p_0}\cdot\vec{p_B})^2/E^{2}_{0}-{p_B}^2}$, 
and the energy difference $\de \equiv E^{*}_B-E^*_{0}/2$, 
where $E$ and $p$ are energy and momentum, 
the subscripts $B$ and $0$ refer to the candidate $B$ and to the 
$e^+e^-$ system respectively and the asterisk denotes the 
$e^+e^-$ CM frame.  For signal events, $\mes$ is centered around the $B$ mass with a 
   resolution of about 2.5 \mevcc, and $\Delta E$ is centered at zero with
a resolution of 12.5 \mev.  
The $B$ candidates are required to have $\Delta E$ in the range $[-0.025,0.025]$ GeV. 
As it will be explained in Sec. \ref{sec:MLfit}, the variable \mes is used in the fit procedure for the signal 
extraction. For this reason, the requirement on it are quite loose: \mes $\in [5.20,5.29]$ \gevcc. 
The region $5.20\gevcc <\mes< 5.27\gevcc$, free from any signal contribution, is 
exploited in the fit to characterize the background directly on data. 
The proper time interval $\deltat$ between the two $B$ decays is calculated from the measured
separation, $\deltaz$, between the decay points of the
reconstructed $B$ ($B_{\rm rec}$) and the other $B$ ($B_{\rm oth}$)
along the beam direction. We accept events with calculated $\deltat$ uncertainty 
less than 2.5 ps and $|\deltat|<$20 ps. 
In less than $1\%$ of the cases, multiple candidates are 
present in the same event and 
we choose the one with 
reconstructed $\Dtilde^0$ mass closest to the nominal mass \cite{ref:PDG}.
In the case of two $B$ candidates reconstructed from the 
same $\Dtilde^0$, we choose the candidate with the largest 
value of $|\cos\theta_{Hel}|$.
The overall reconstruction and selection efficiency for signal, evaluated on 
MC, is $(10.8 \pm 0.5) \%$.

\section{Background characterization}
After applying the selection criteria described above, the 
background is composed of continuum events ($e^+ e^- \rightarrow q\bar{q}$, $q=u,d,s,c$) 
and $\Upsilon(4\mbox{S}) \rightarrow \bb$ events (``$\bb$'', in the following).
To discriminate against the continuum background events (the dominant background component), 
which, in contrast to $\bb$ events, have a jet-like shape, we use a 
Fisher discriminant \fis\ \cite{ref:Fish}. The discriminant \fis\ is a linear combination 
of three variables: $\cos\theta_{thrust}$, the cosine of the angle between 
the $B$ thrust axis and the thrust axis of the rest of the event, $L_0=\sum_{i} p_i$, and 
$L_2 =\sum_{i} p_i  |\cos \theta_i|^2$. Here, $p_i$ is the momentum and $\theta_i$ 
is the angle with respect to the thrust axis of the $B$ candidate. The index $i$ runs 
over all the reconstructed tracks and energy deposits in the calorimeter not associated with 
a track. The tracks and energy deposits used to reconstruct the $B$ are excluded from these 
sums. 
All these variables are calculated in the 
$e^+e^-$ CM frame. The coefficients of the Fisher discriminant, 
chosen to maximize the separation between signal and continuum background, are determined 
using signal MC events and off-resonance data. 
A cut on this variable with 85$\%$ efficiency on simulated signal events 
would reject about 80$\%$ of continuum background events, as estimated on 
off-resonance data. We choose not to cut on the Fisher discriminant, as we 
will use this variable in the fit procedure to extract the signal.  
The variable $\deltat$ gives further discrimination between signal and 
continuum events. 
For events in which the $B$ meson has been correctly reconstructed, the $\deltat$ distribution 
is the convolution of a decreasing exponential function $e^{-t/\tau_B}$ (with $\tau_B$ equal to the 
$B$ lifetime) with the 
resolution on $\Delta z$ from the detector reconstruction.
The distribution is then wider than in the case of continuum events, in which 
just the resolution effect is observed.

The  $B_{\rm rec}$ decay point is the common vertex of the $B$ decay 
products. The  $B_{\rm oth}$ decay point is obtained using tracks which 
do not belong to $B_{\rm rec}$ and imposing constraints from the 
$B_{\rm rec}$ momentum and the beam-spot location.
 
Background events for which the reconstructed $\KS$, $\pi^+$, and 
$\pi^-$ come from a real $\Dtilde^0$ (``true $D^0$'', in the following) 
are treated separately because of their distribution over the $\Dtilde^0$ 
Dalitz plane.
A fit to the $K_S\pi^+\pi^-$ invariant mass distribution for events in 
the \mes\ sideband ($\mes < 5.27 \gevcc$) has been performed on data to obtain 
the fraction of true \Dz\ equal to 0.289 $\pm$ 0.028. 
This value is in agreement with that determined from simulated $\bb$ and continuum 
background samples. 

Background events with final states containing ${\Dz} h^+\pi^-$  or ${\Dzb} h^-\pi^+$,
where $h^{\pm}$ is a candidate $K^{\pm}$ and $\Dtilde^0\to K_S\pi^+\pi^-$, can mimic 
$b\to u$ mediated signal events 
(see Fig. \ref{fig:feyn}). 
The fraction of these events (relative to the number of true $D^0$ events), defined as
$R_{b\to u} = \frac{N ({\Dz} h^+\pi^-) + N ({\Dzb} h^-\pi^+)}{N ({\Dz}h^+\pi^-) + N ({\Dz}h^-\pi^+) + N ({\Dzb}h^+\pi^-) + N ({\Dzb}h^-\pi^+)}$, 
has been found to be $0.88\pm 0.02$ and $0.45\pm 0.12$ in $\bb$ and 
continuum MC events respectively.

Studies have been performed on $B$ decays, which have the same final state 
reconstructed particles as the signal decay (so called peaking background).  
From MC studies, we identify three possible background sources of this kind: 
$\Bz\rightarrow \Dtilde^0 \Kstarz$ ($\Kstarz\rightarrow K^{+}\pi^{-}$, 
$\Dtilde^0 \rightarrow \pi^{+}\pi^{-}\pi^{+}\pi^{-}$), $\Bz\rightarrow \Dtilde^0 \rho^{0}$ 
($\rho^{0}\rightarrow \pi^{+}\pi^{-}$, $\Dzb \rightarrow \KS\pi^{+}\pi^{-}$, 
where $\rho^0$ is reconstructed as a $\Kstarz$ with a misidentified pion) and 
charmless events of the kind $\Bz\rightarrow \Kstarz \KS\KS$.  
To precisely evaluate the selection efficiency for $\Dtilde^0 \rho^0$ and 
$\Dtilde^0 \Kstarz$ with $\Dtilde^0 \rightarrow \pi^{+} \pi^{-} \pi^{+} \pi^{-}$, 
dedicated MC samples have been generated, resulting in 
$(0.04 \pm 0.02) \%$ or $(0.18 \pm 0.04) \%$, respectively.  
With these efficiencies, we expect to select about 0.9 $\Dtilde^0 \rho^0$ events and 
0.1 $\Dtilde^0 \rightarrow 4\pi$ events in $371\times 10^6$ $\bb$ pairs.
In the latter case the requirement on $\alpha_{\KS}$ rejects most of the background, 
while for $\Dtilde^0 \rho^0$ the cuts on $\Delta E$ and the particle identification 
of the $K^{\pm}$ are the most effective.  
The number of charmless background events has been evaluated on data 
from the $\Dtilde^0$ mass sidebands, namely $M_{K_S\pi^+\pi^-}$ in the 
range [1.810, 1.839] or [1.889, 1.920] $\gevcc$; 
we obtain $N_{peak}=-5\pm 7$ events, consistent with 0.  
Hence we assume these background sources can be neglected in our signal extraction 
procedure; the effects of this assumption are taken into account in the evaluation 
of the systematic uncertainties.  
The remaining $\bb$ background is combinatorial.
\section{Likelihood fit and measured yield}
\label{sec:MLfit}
We perform an unbinned extended maximum likelihood fit to the variables 
$m_{\rm ES}$, $\F$ and $\Delta t$, in order to extract the 
signal, continuum and $\bb$ background yields, probability density 
function (PDF) shape parameters, and \CP\ parameters. 
We write the likelihood as
\bea
{\cal L} & = & \frac{e^{-\eta} \eta^N}{N!}\prod_{\alpha} \prod_{i=1}^{N_\alpha} {\cal P^{\alpha}}(i)~,
\label{eq:like-yields}
\eea
where ${\cal P}^\alpha(i)$ and $N_\alpha$ are the PDF for event $i$ and the total number 
of events for component $\alpha$ (signal, $\bb$ background, continuum background).
Here $N$ is the total number of selected events and $\eta$ is the expected 
value for the total number of events, according to Poisson statistics. 
The PDF is the product of a ``yield'' 
PDF ${\cal P}^\alpha(\mes) {\cal P}^\alpha(\F) {\cal P}^\alpha(\Delta t)$ 
(written as a product of 1D PDFs since $m_{\rm ES}$, $\F$ and $\Delta t$ 
are not correlated) and of the $D^0$ Dalitz plot dependent part: 
${\cal P}^\alpha(m_+^2,m_-^2)$ (where $m_+^2=m^2_{K_S\pi^+}$ and 
$m_-^2=m^2_{K_S\pi^-}$).

The $m_{\rm ES}$ distribution is parametrized by a Gaussian function 
for the signal and by an Argus function \cite{argus} that is
different for continuum and $\bb$ backgrounds. The \F\ distribution 
is parametrized using a asymmetric Gaussian distribution for the signal and $\bb$ 
background and the sum of two Gaussian distributions for the continuum background. 
For the signal, $|\Delta t|$ is parametrized with an exponential decay PDF $e^{-t/\tau}$ 
in which $\tau = \tau_{\Bz}$ \cite{ref:PDG}, convolved with a resolution 
function that is a sum of three Gaussians \cite{ref:s2b}.
A similar parametrization is used for the backgrounds using exponential 
distributions with effective lifetimes.

The continuum background parameters are obtained from off-resonance data, while the $\bb$ 
parameters are taken from MC. 
The fractions of true $D^0$ and the ratios $R_{b\to u}$ in the backgrounds are 
fixed in the fit to the values obtained on data and MC respectively.

Using the effective parameters defined in Eq. \ref{eq:rs_square} and \ref{eq:k}, the partial decay rate 
for events with true \Dz\ can be written as follows: 
\begin{eqnarray}
&&\Gamma(\Bz\to D[K_S\pi^-\pi^+]K^+\pi^-)  \propto  |{\cal P}^{\sig}_{-}|^2 + 
 \nonumber \\
&&r_S^2|{\cal P}^{\sig}_{+}|^2 + 2kr_S|{\cal P}^{\sig}_{-}||{\cal P}^{\sig}_{+}|\cos(\delta_S+\delta_{D-} - \gamma)\, ,\\
&&\Gamma(\overline{\Bz}\to D[K_S\pi^-\pi^+]K^-\pi^+)  \propto  |{\cal P}^{\sig}_{+}|^2 + 
 \nonumber \\
&&r_S^2|{\cal P}^{\sig}_{-}|^2 + 2kr_S|{\cal P}^{\sig}_{+}||{\cal P}^{\sig}_{-}|\cos(\delta_S+\delta_{D+} + \gamma)\, ,
\label{eq:dalitzrate1}
\end{eqnarray}
where ${\cal P}^{\sig}_{+} \equiv {\cal P}^{\sig}(m_+^2,m_-^2)$, 
${\cal P}^{\sig}_{-} \equiv {\cal P}^{\sig}(m_-^2,m_+^2)$ and where 
$\delta_{D +} \equiv \delta_D(m_+^2,m_-^2)$ is the strong phase difference between 
${\cal P}^{\sig}_{+}$ and ${\cal P}^{\sig}_{-}$ and $\delta_{D -} \equiv \delta_D(m_-^2,m_+^2)$ 
is the strong phase difference between ${\cal P}^{\sig}_{-}$ and ${\cal P}^{\sig}_{+}$.  

For the resonance structure of the $D^0\to K_S\pi^+\pi^-$ decay amplitude, 
${\cal P}^{\sig}_{+}$, we use the same model as documented in 
\cite{ref:DKDalitzbabar}.  
This is determined on a large data sample (about 487000 events, with 97.7$\%$ purity) from a Dalitz plot analysis of 
$D^0$ mesons from $D^{*+}\to D^0\pi^+$ decays produced in $e^+e^- \to c\bar{c}$ events.  
The decay amplitude is parametrized, using an isobar model, with the sum of the contributions of ten two-body
decay modes with intermediate resonances. In addition, the K-matrix approach \cite{ref:Kmatrix} is used to 
describe the S-wave component of the $\pi^+\pi^-$ system, which is characterized by the overlap of broad 
resonances. 
The systematic effects of the assumptions made on the model used to describe the decay 
amplitude of neutral $D$ mesons into $K_S\pi^+\pi^-$ final states are evaluated, as it 
will be described. 
To account for possible selection efficiency variations across the 
Dalitz plane, the efficiency is parametrized with a polynomial function 
whose parameters are evaluated on MC.  
This function is convoluted with the Dalitz distribution ${\cal P}^{\sig}_{+}$.  
The distribution over the Dalitz plot for events with no true \Dz\ is parametrized 
with a polynomial function whose parameters are evaluated on MC.  

Following Ref.~\cite{ref:stephane}, we have performed a study to evaluate the possible 
variations of $r_S$ and $k$ over 
the $B^0\to\Dtilde^0 K^+ \pi^-$ Dalitz plot. For this purpose 
we have built a \Bz\ Dalitz model suggested by recent measurements \cite{ref:cavoto, ref:polci}, 
including $K^{*}(892)^0$, $K_0^{*}(1430)^0$, $K_2^{*}(1430)^0$, $K^{*}(1680)^0$, 
$D_{s2}(2573)^{\pm}$, $D_{2}^{*}(2460)^{\pm}$, and $D_{0}^{*}(2308)^{\pm}$ contributions. 
We have considered the region within 48 \mevcc of the nominal mass of the $K^{*}(892)^0$ 
resonance and obtained the distribution of $r_S$ and $k$ by randomly varying 
all the strong phases ([0,2$\pi$]) and the amplitudes (within [0.7,1.3] of their nominal value). 
The ratio between $b\to u$ and $b\to c$ amplitudes for 
each resonance has been 
fixed to 0.4. In the $K^{*}(892)^0$ mass region, we find that $r_S$ varies between 
0.30 and 0.45 depending upon the values of the contributing phases and of the amplitudes. 
The distribution of $k$ is quite narrow, centered at 0.95 with an r.m.s. of
0.02. The study has been repeated varying the ratio between $b\to u$ and $b\to c$ 
amplitudes between 0.2 and 0.6, leading to very similar results.  
For these reasons the value of $k$ has been fixed to 0.95 and a variation of 0.03 
has been considered for the systematic uncertainties evaluation. 
On the contrary, $r_S$ will be extracted from data.

\begin{figure*}[htb] 
\begin{center}   
\epsfig{figure=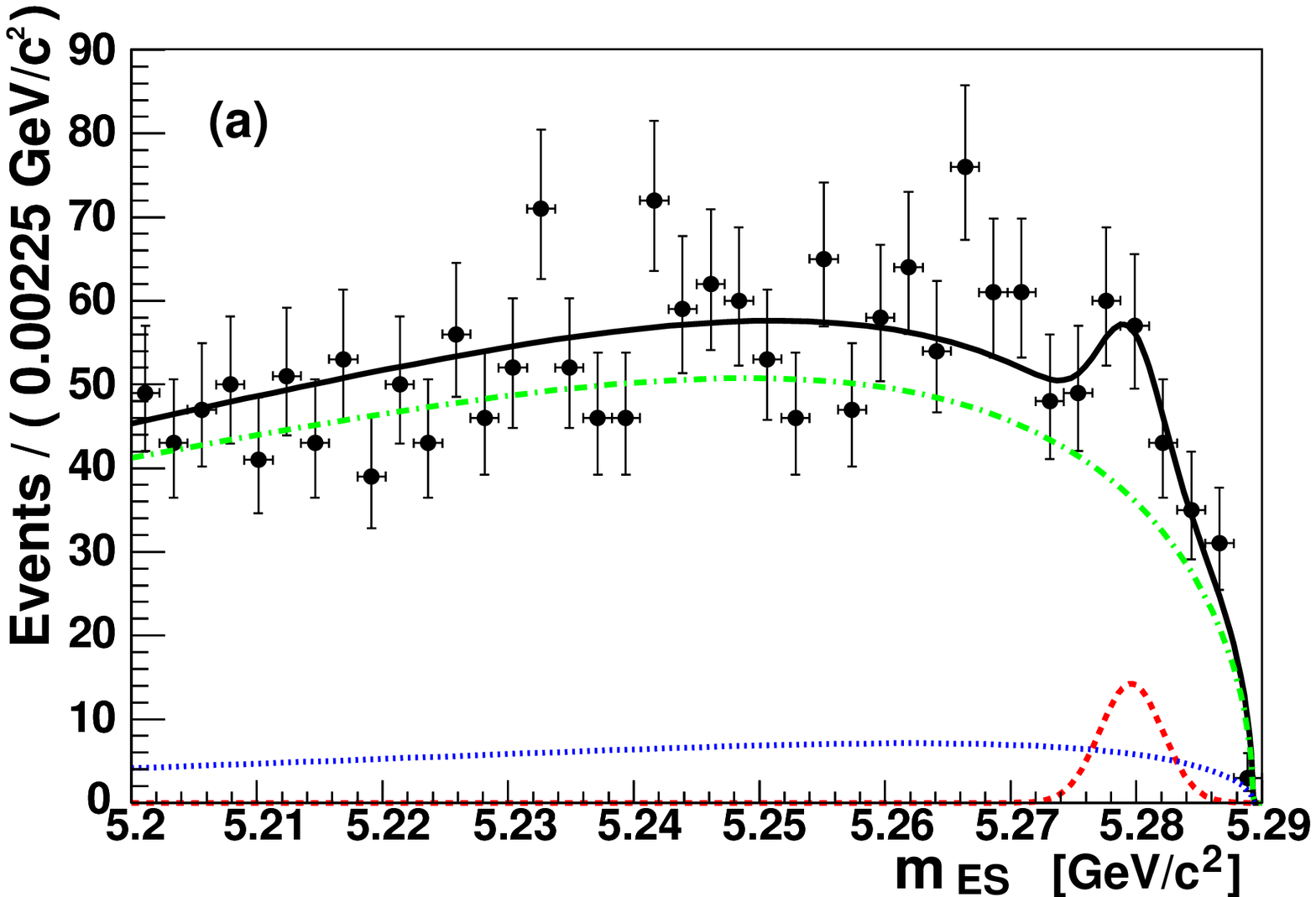,width=8.cm}
\hspace{0.5truecm}
\epsfig{figure=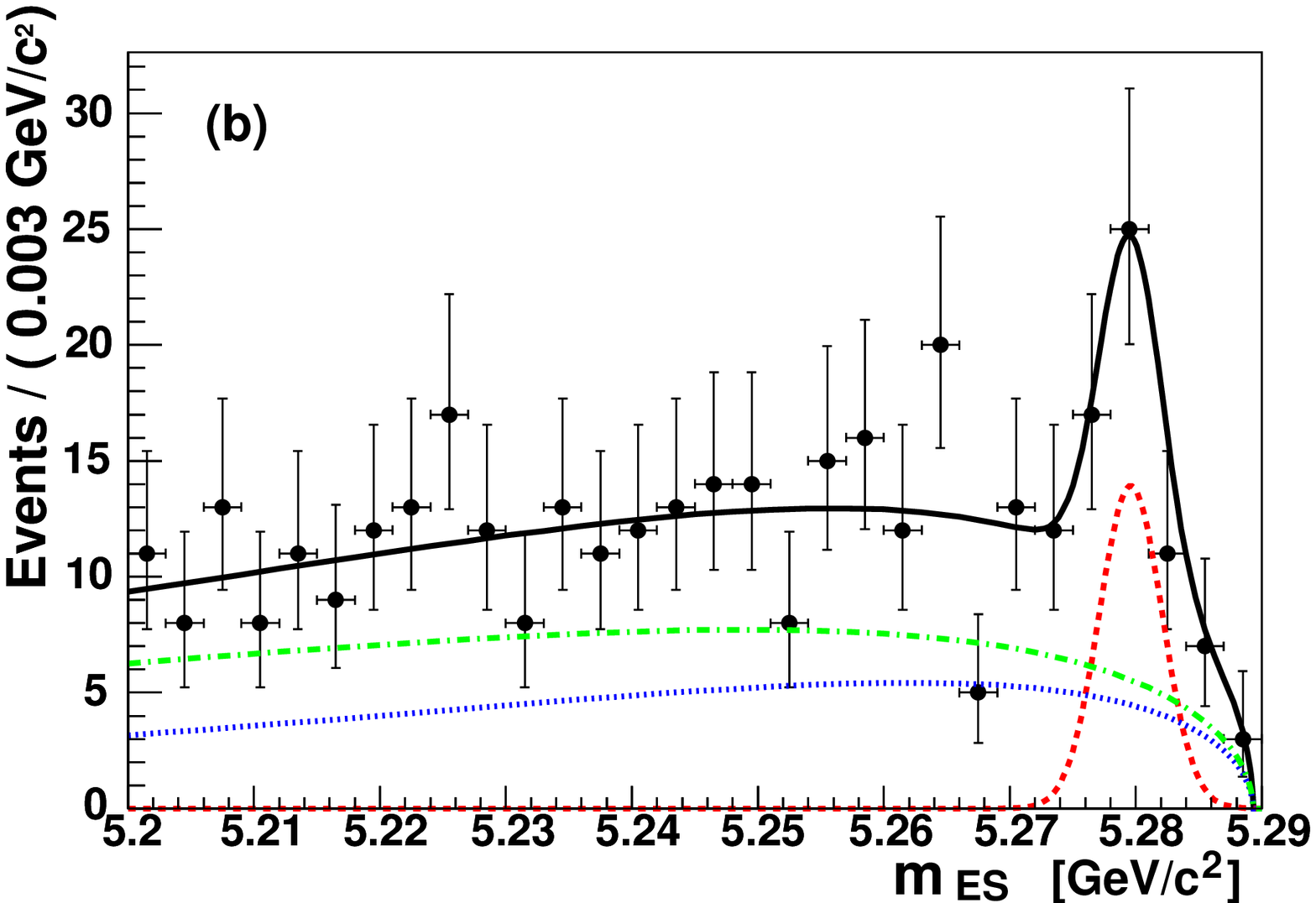,width=8.cm}\\
\epsfig{figure=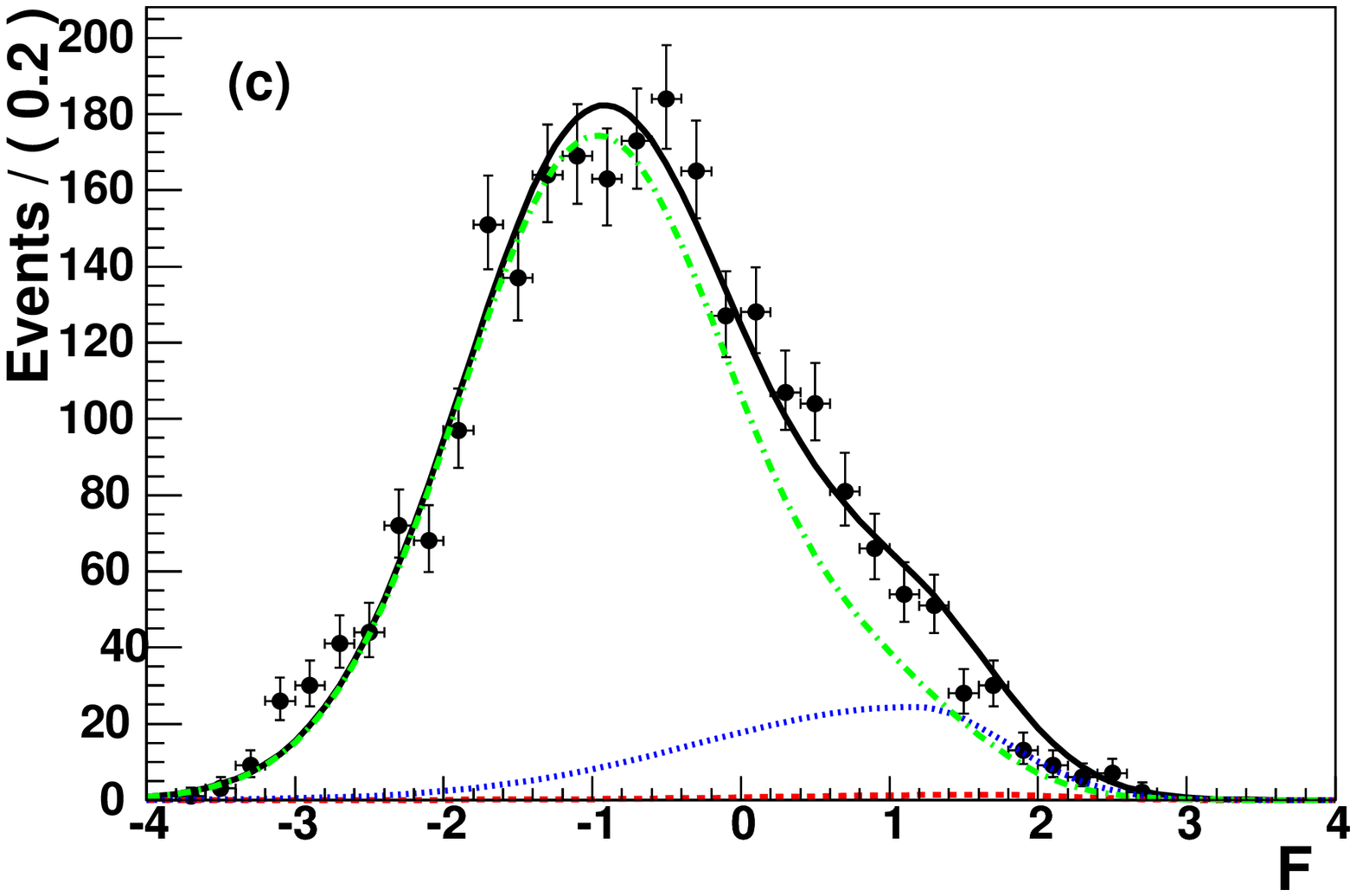,width=8.cm}
\hspace{0.5truecm}
\epsfig{figure=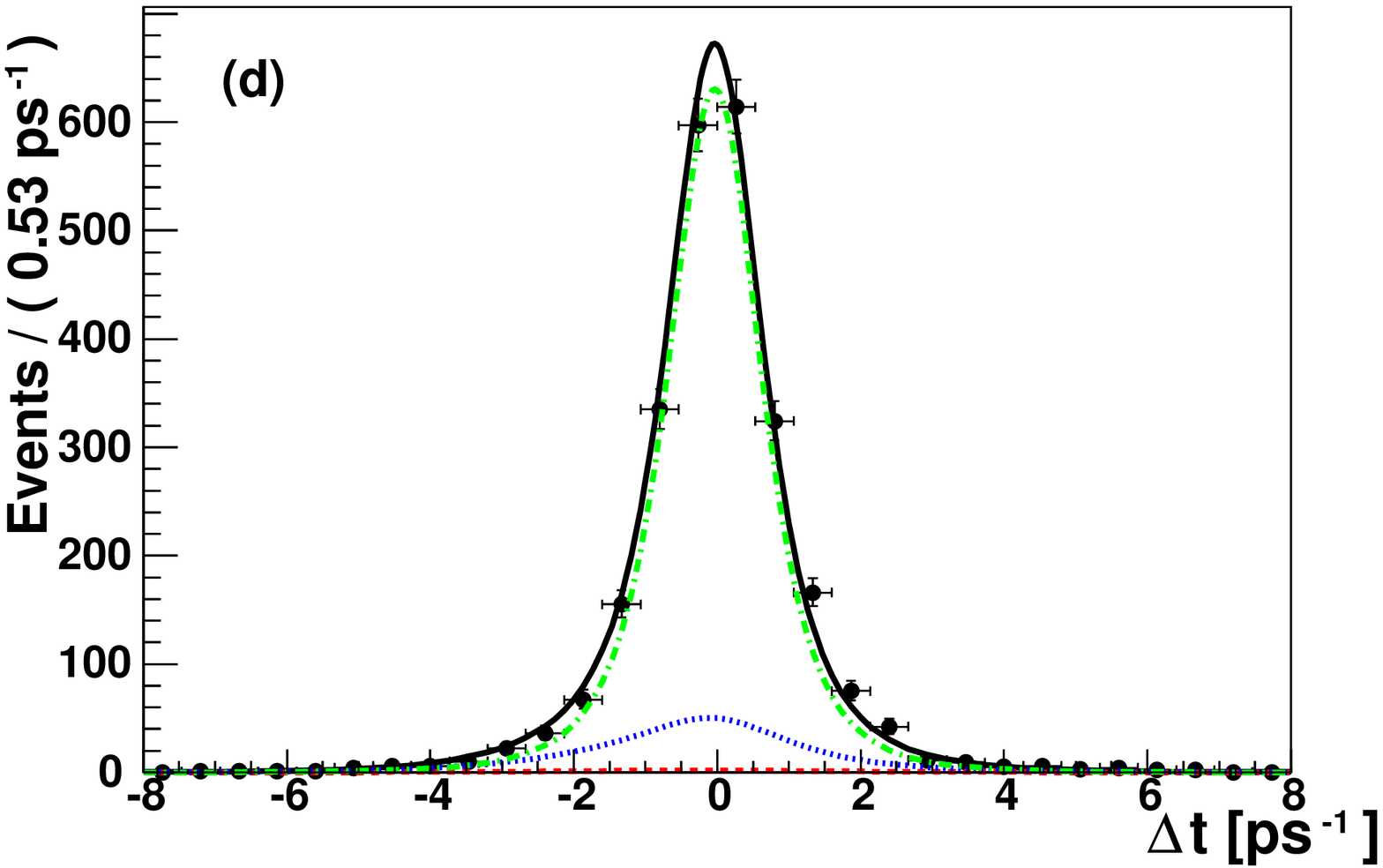,width=8.cm}
\end {center} 
\caption{ \mes\ projection from the fit (a).
The data are indicated with dots and error bars and the different 
fit components are shown: signal (dashed), $\bb$ (dotted) and 
continuum (dot-dashed). With a different binning (b), 
\mes\ projection after a cut on $\fis >0.4$ is applied, to 
visually enhance the signal. $\fis$ and $\Delta t$ projection from the 
fit (c, d). \label{fig:shapes_dkst}}
\end{figure*}

We perform the fit for the yields on data extracting the number of events 
for signal, continuum and $\bb$, as well as the slope of the Argus 
function for the $\bb$ background. 
The fitting procedure has been validated using simulated events. 
We find no bias on the number of fitted events for any of the components.
The fit projection for \mes is shown in Fig. \ref{fig:shapes_dkst}. 
We find $39 \pm 9$ signal, $231\pm 28$ $\bb$ and $1772\pm 48$ continuum 
events.
In Fig. \ref{fig:shapes_dkst} we also show, for illustration purposes, 
the fit projection for \mes, after a cut on $\fis >0.4$ is applied, to 
visually enhance the signal.  Such a cut has an approximate efficiency of 75$\%$ on 
signal, while it rejects 90$\%$ of the continuum background. 

\section{Determination of $\gamma$}
From the fit to the data we obtain a three-dimensional likelihood $\mathcal{L}$ for 
$\gamma$, $\delta_S$ and $r_S$ which includes only statistical uncertainties.  
We convolve this likelihood with a three-dimensional Gaussian that takes into 
account the systematic effects, described later, in order to obtain the 
experimental three-dimensional likelihood for $\gamma$, $\delta_S$ and $r_S$.  
From simulation studies we observe that, due to the small signal statistics and 
the high background level, $r_S$ is overestimated and the error on $\gamma$ is underestimated, 
when we project the experimental three-dimensional likelihood on either $r_S$ or $\gamma$, 
after integrating over the other two variables. 
This problem disappears if either $r_S$ is fixed in the fit or if we combine the 
three-dimensional likelihood function ($\gamma,\delta_S$,$r_S$) obtained from this 
data sample with external information on $r_S$. In the following we will show the results 
of both these approaches.

The systematic uncertainties, summarized in Table \ref{tab:syst}, 
are evaluated separately on $\gamma,\delta_S$ and $r_S$ and considered uncorrelated 
and Gaussian.  It can be noted that the systematic error is much 
smaller than the statistical one.  
The systematic uncertainty from the Dalitz model used to describe true 
$D^0\to K_S\pi^+\pi^-$ decays is evaluated on data by repeating the fit with 
models alternative to the nominal one, as described in detail in \cite{ref:DKDalitzbabar}.  
The $D^0\to K_S\pi^+\pi^-$ Dalitz model is known to be the source of the 
largest systematic contribution in this kind of measurements \cite{ref:DKDalitzbabar,ref:BelleDalitz}. 
All the other contributions have been evaluated on a high statistics simulated 
sample in order not to include statistical effects.  
To evaluate the contribution related to $m_{\rm ES}$, \F\ and $\Delta t$ PDFs, 
we repeat the fit by varying the PDF parameters obtained from MC within 
their statistical errors. 
To evaluate the uncertainty arising from the assumption of negligible 
peaking background contributions, the true \Dz\ fraction and $R_{b\to u}$ 
in the background, we repeat the fit by varying the 
number of these events and fractions within their statistical errors. 
The uncertainty from the assumptions on the factor $k$ is also evaluated.  
The reconstruction efficiency across the Dalitz plane for true \Dz\ events 
and the Dalitz plot distributions for background with no true \Dz\ have 
been parametrized on MC using polynomial functions.  Systematics uncertainties 
have been evaluated by repeating the fit assuming the efficiency and the 
distribution for these backgrounds to be flat across the Dalitz plane.  

\begin{figure}[!ht]
\begin{center}   
\epsfig{figure=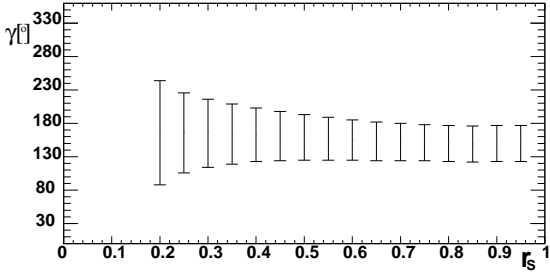,width=8.cm}
\caption{ $68\%$ probability regions obtained for $\gamma$, for different values of 
$r_S$. For values of $r_S$ lower than 0.2, the distribution obtained for 
$\gamma$ is almost flat and hence does not allow to determine 
significative $68\%$ probability regions. 
The solution corresponding to a 180$^{\circ}$ ambiguity is not shown. 
\label{fig:Gamma_fixedrB}}
\end{center}
\end{figure}

\begin{table}[htpb]
\begin{center}
\begin{tabular}{c||c|c|c}
\hline
Systematics source            &$\Delta\gamma [{}^{o}]$ & $\Delta\delta_S [{}^{o}]$&  $\Delta r_S (10^{-2})$  \\ \hline
Dalitz model for signal       &    6.50   &  15.80   &  6.00  \\         
PDF shapes                    &    1.50   &  2.50    &  5.20  \\
Peaking background            &    0.14  &  0.12   &  0.04 \\
$k$ parameter                 &    0.07  &  1.20    &  7.10  \\
True \Dz\ in the background  &    0.05  &  0.03   &  1.00  \\
$R_{b\to u}$         &    0.01  &  1.10    &  1.90  \\
Efficiency variation             &    0.31  &  0.62   &  0.61 \\
Dalitz background param.      &    0.03  &  0.27   &  0.20  \\\hline 
Total                         &    6.70   &  16.10   &  11   \\ \hline         
\end{tabular}
\end{center}
\caption{Systematics uncertainties on $\gamma$ , $\delta_S$ , and $r_S$. \label{tab:syst}}
\end{table}
In Fig. \ref{fig:Gamma_fixedrB}, we show the $68\%$ probability region
obtained for $\gamma$ assuming different fixed values of $r_S$ and 
integrating over $\delta_S$. For values of $r_{S}<0.2$ we do not have a 
significant measurement of $\gamma$.  
The value of (the fixed) $r_S$ does not affect the central value of $\gamma$, 
but its error. For example, for $r_S$ fixed to $0.3$, we obtain 
$\gamma = (162 \pm 51)^{\circ}$. On MC, for the same fit configuration, 
the average error is $45^{\circ}$ with a r.m.s. of $14^{\circ}$.  
The \babar\ analysis for charged $B$ decays \cite{ref:DKDalitzbabar}, 
using the same Dalitz technique for $\Dtilde^0 \rightarrow \KS\pi^{+}\pi^{-}$, 
gives, for a similar luminosity, an error 
on $\gamma$ of $29^{\circ}$, from the combination of 
$B^\pm\to \Dtilde^0K^{\pm}$, $B^{\pm}\to \Dtilde^{*0}K^{\pm}$ and 
$B^{\pm}\to \Dtilde^0K^{*\pm}$.  
The use of neutral $B$ decays can hence give a contribution to the improvement 
of the precision on $\gamma$ determination comparable with that of a single charged 
$B$ channel.  

\begin{figure*}[!ht]
\begin{center}   
\epsfig{figure=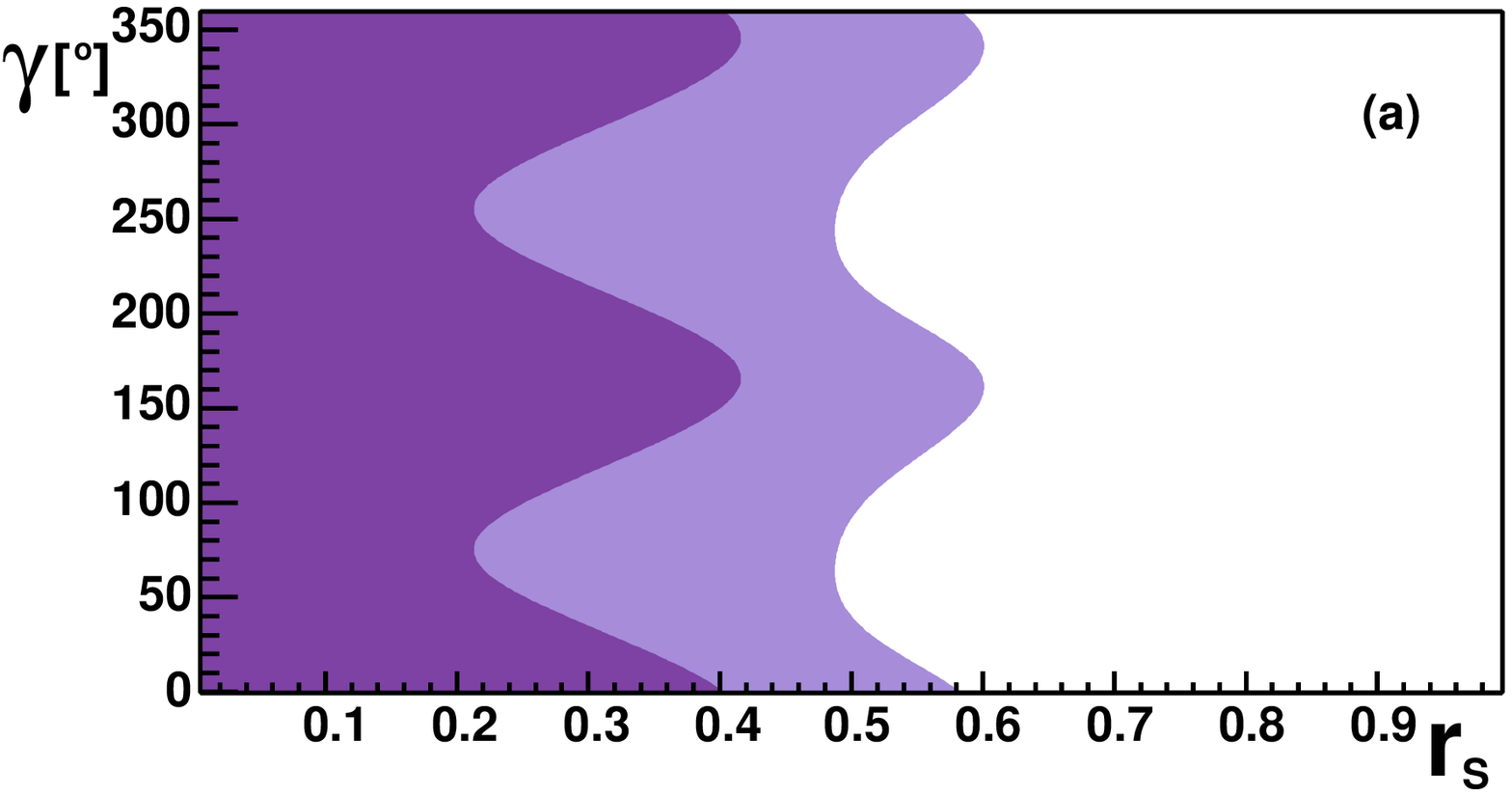,width=8.cm, height=4.75cm}
\hspace{-0.6truecm}
\epsfig{figure=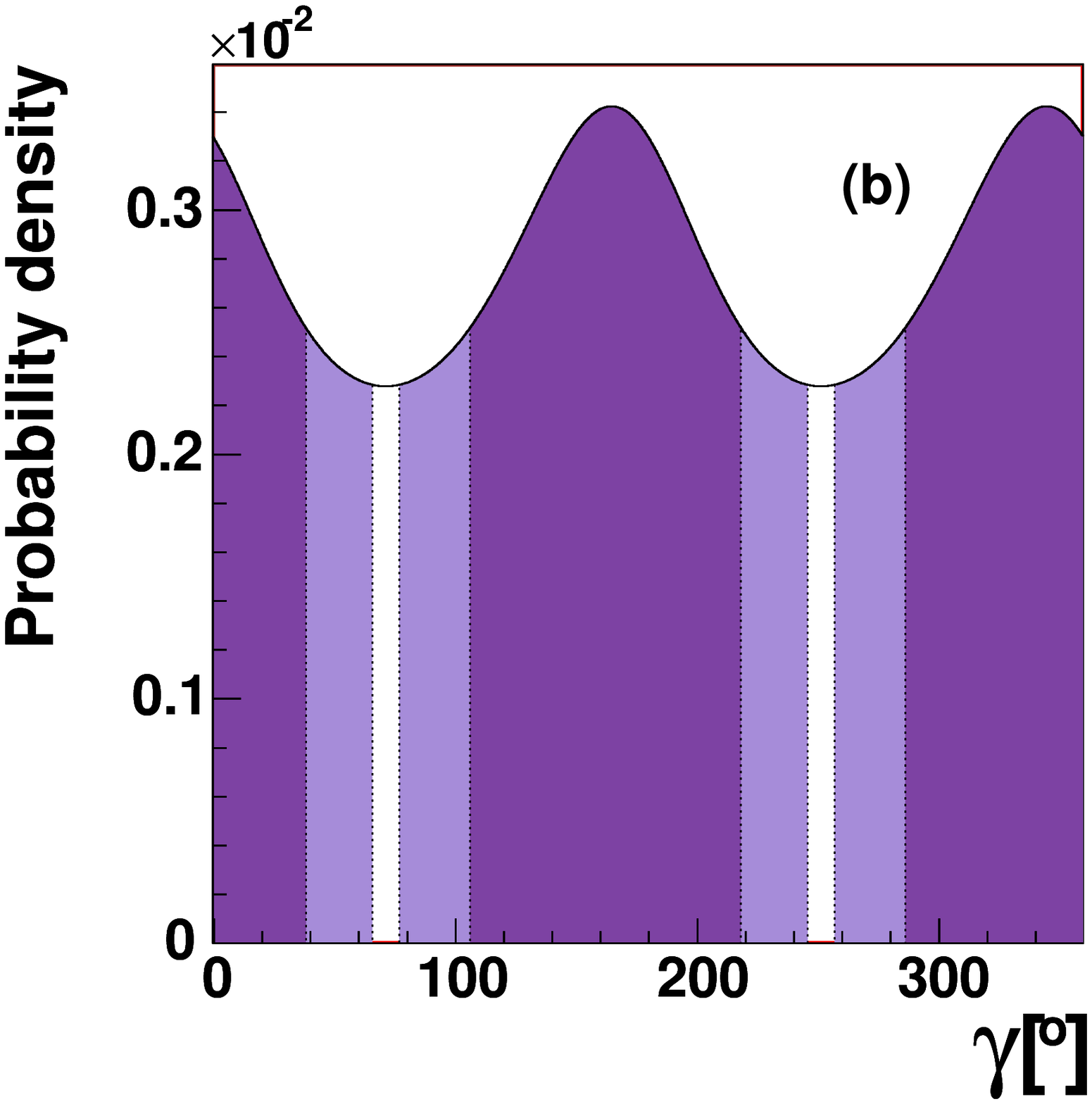,width=5.0cm}
\hspace{-0.4truecm}
\epsfig{figure=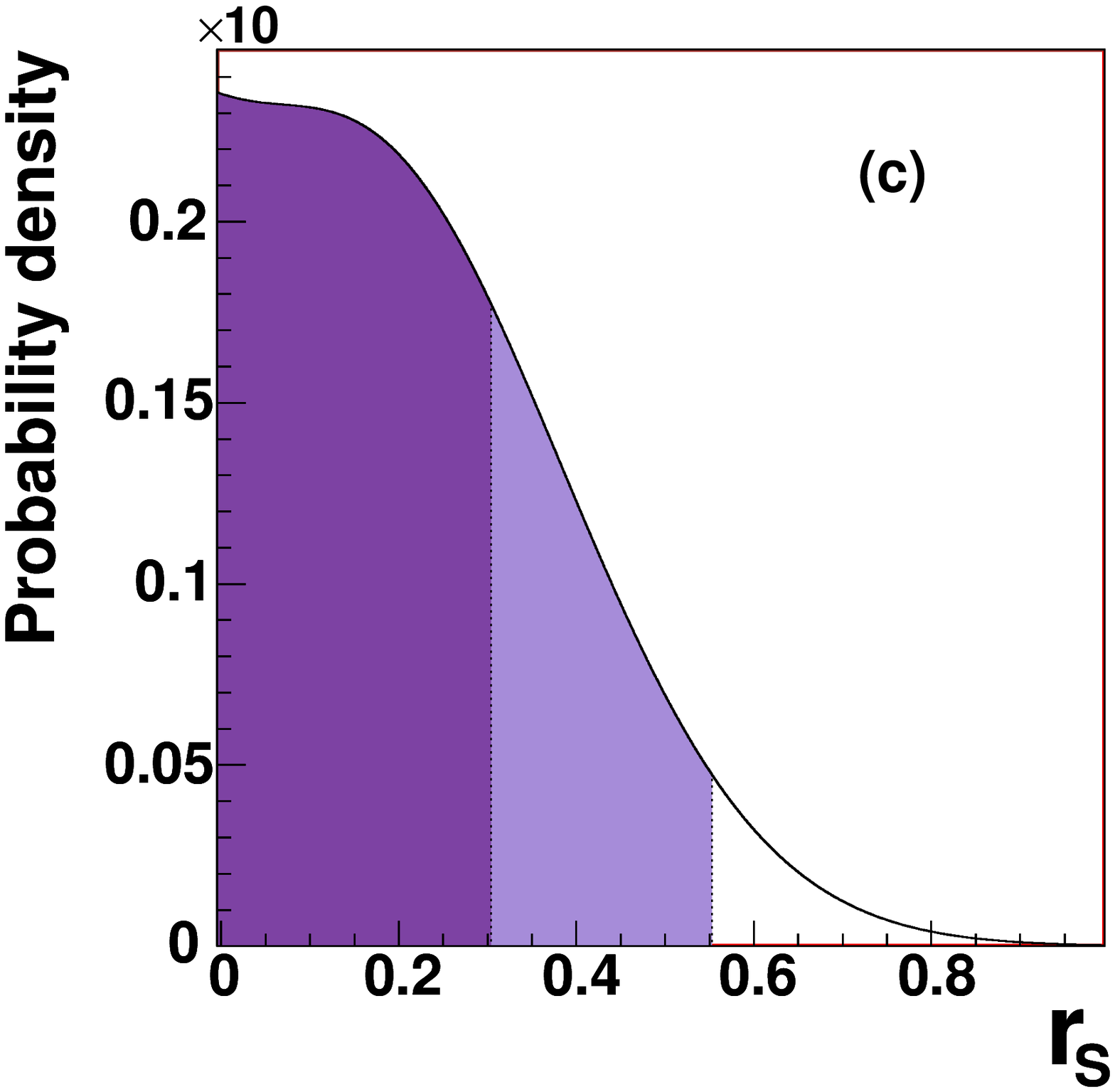,width=5.0cm}
\caption{68 \% (dark shaded zone) and 95 \% (light shaded zone) probability 
regions for the combined PDF projection on the $\gamma$ vs $r_S$ plane 
(a), $\gamma$ (b) and $r_S$ (c).
\label{fig:gammavsrb_data_comb}}
\end{center}
\end{figure*}

Combining the final three-dimensional PDF with the PDF for 
$r_S$ measured with an ADS method \cite{ref:ADS}, reconstructing the neutral 
$D$ mesons into flavor modes \cite{ref:sha}, we obtain, at 68$\%$ probability, 
\begin{eqnarray}
&\gamma&= (162 \pm 56)^{\circ} \mbox{ or }(342 \pm 56)^{\circ};\label{eq:gamma68} \\\  
&\delta_S&= (62 \pm 57)^{\circ} \mbox{ or }(242 \pm 57)^{\circ};\label{eq:delta68}\\ 
&r_S&  < 0.30\label{eq:rS68};
\end{eqnarray}
while, at 95$\%$ probability:
\begin{eqnarray}
&\gamma& \in [77, 247]^{\circ} \mbox{ or } [257, 426]^{\circ};\label{eq:gamma95} \\
&\delta_S& \in [-23, 147]^{\circ} \mbox{ or } [157, 327]^{\circ};\label{eq:delta95}\\ 
&r_S&  < 0.55\label{eq:rS95}.
\end{eqnarray}

The preferred value for $\gamma$ is somewhat far from the value obtained 
using charged $B$ decays, which is around 75$^o$ for both \babar\ and 
Belle Dalitz analyses, but is compatible with both the results within about 1.5$\sigma$.  
In Fig. \ref{fig:gammavsrb_data_comb} we show the distributions we obtain 
for $\gamma$, $r_S$ and $\gamma$ vs. $r_S$ (the 68\% and 95\% probability 
regions are shown in dark and light shading respectively).
The one-dimensional distribution for a single variable is obtained from the 
three-dimensional PDF by projecting out the variable and integrating over the others.
\section{Conclusions}
In summary, we have presented a novel technique for extracting the angle 
$\gamma$ of the Unitarity Triangle in $\Bz \rightarrow \Dtilde^0 \Kstarz$ 
($\Bzb \rightarrow \Dtilde^0 \Kstarzb$)with the $\Kstarz \rightarrow K^{+}\pi^{-}$ 
($\Kstarzb \rightarrow K^{-} \pi^{+}$), using a Dalitz analysis of 
$\Dtilde^0 \rightarrow \KS\pi^{+}\pi^{-}$. 
With the present data sample, interesting results on $\gamma$ (Eqs. \ref{eq:gamma68}, \ref{eq:gamma95}) 
and $r_S$ (Eqs. \ref{eq:rS68}, \ref{eq:rS95}) are 
obtained when combined with the determination of $r_S$ from the study of $\Dtilde^0$
decays into flavor modes.  
The result for $\gamma$ is consistent, within 1.5$\sigma$, with the determination 
obtained using charged $B$ mesons.
If the ratio $r_S$ is found to be of the order of $0.3$, the use of 
neutral $B$ mesons, proposed here, could give results on $\gamma$ as precise 
as those obtained using similar techniques and charged $B$ 
mesons \cite{ref:DKDalitzbabar}. 

We are grateful for the 
extraordinary contributions of our \pep2\ colleagues in
achieving the excellent luminosity and machine conditions
that have made this work possible.
The success of this project also relies critically on the 
expertise and dedication of the computing organizations that 
support \babar.
The collaborating institutions wish to thank 
SLAC for its support and the kind hospitality extended to them. 
This work is supported by the
US Department of Energy
and National Science Foundation, the
Natural Sciences and Engineering Research Council (Canada),
the Commissariat \`a l'Energie Atomique and
Institut National de Physique Nucl\'eaire et de Physique des Particules
(France), the
Bundesministerium f\"ur Bildung und Forschung and
Deutsche Forschungsgemeinschaft
(Germany), the
Istituto Nazionale di Fisica Nucleare (Italy),
the Foundation for Fundamental Research on Matter (The Netherlands),
the Research Council of Norway, the
Ministry of Education and Science of the Russian Federation, 
Ministerio de Educaci\'on y Ciencia (Spain), and the
Science and Technology Facilities Council (United Kingdom).
Individuals have received support from 
the Marie-Curie IEF program (European Union) and
the A. P. Sloan Foundation.



\begin{thebibliography}{99}
\bibitem{ref:GLW}         M. Gronau and D. London,  Phys. Lett. {\bf B253}, 483 (1991);~
                          M. Gronau and D. Wyler,  Phys. Lett. {\bf B265}, 172 (1991);~
                          I. Dunietz,  Phys. Lett. {\bf B270}, 75 (1991);~
                          I. Dunietz,  Z. Phys. {\bf C56}, 129 (1992).
   \bibitem{ref:ADS} D.~Atwood, I.~Dunietz, and A.~Soni, Phys. Rev. Lett. {\bf 78}, 3257 (1997);~
                                                         Phys. Rev. {\bf D63}, 036005 (2001).
\bibitem{ref:DKDalitz}    
                          A. Giri, Yu. Grossman, A. Soffer, and J. Zupan, Phys. Rev. {\bf D68}, 054018 (2003).
\bibitem{ref:ckm}  N. Cabibbo, Phys. Rev. Lett. {\bf 10} (1963) 531;
                   M. Kobayashi and T. Maskawa, Prog. Theor. Phys. {\bf 49} (1973) 652.
\bibitem {ref:DKDalitzbabar}
  B.~Aubert {\it et al.}  (BABAR Collaboration), Phys.\ Rev.\  D {\bf 78}, 034023 (2008).
\bibitem{ref:BelleDalitz}
  K.~Abe {\it et al.}  (Belle Collaboration),
  arXiv:0803.3375 [hep-ex].
\bibitem{gronau2002} M.~Gronau, Phys. Lett. {\bf B557}, 198 (2003).
\bibitem{Bona:2005vz}     M.~Bona {\it et al.}  (UTfit Collaboration),  JHEP {\bf 0507}, 028 (2005)  [arXiv:hep-ph/0501199]. Updated results available at {\tt http://www.utfit.org/}.
\bibitem{ref:onlyforfairness} J.~Charles {\it et al.} (CKMfitter Collaboration),
Eur. Phys. J. C41, 1 (2005). Updated results available at {\tt  http://ckmfitter.in2p3.fr}. 
\bibitem{ref:sha}         B.~Aubert {\it et al.} (BABAR Collaboration),  Phys. Rev. {\bf D74}, 031101 (2006).
\bibitem{ref:cavoto}       G.~Cavoto {\it et al.}, Proceedings of the CKM 2005 Workshop (WG5), 
                           UC San Diego, 15-18 March 2005 [arXiv:hep-ph/0603019].
\bibitem{ref:det}       B. Aubert {\it et al.} (BABAR Collaboration), Nucl. Instr. and Methods {\bf A479}, 1 (2002).
\bibitem{ref:PDG} C.~Amsler {\it et al.} (Particle Data Group), Phys. Lett. {\bf B 667}, 1 (2008)
\bibitem{ref:Fish}
R.~A.~Fisher, Annals Eugen. {\bf 7}, 179 (1936).
\bibitem{argus}
H.~Albrecht {\em et al.} (ARGUS Collaboration), Z. Phys. {\bf C48}, 543 
(1990).
\bibitem{ref:s2b}  B.~Aubert {\it et al.}  (BABAR Collaboration), Phys.\ Rev.\ Lett.\  {\bf 99}, 171803 (2007) [arXiv:hep-ex/0703021].
\bibitem{ref:Kmatrix} E.~P.~Wigner, \pr{70}, 15 (1946); 
                      S.~U.~Chung {\it et al.}, \annp{4}, 404 (1995).  
\bibitem{ref:stephane}  S. Pruvot, M.-H. Schune, V. Sordini, and A. Stocchi [arXiv:hep-ph/0703292], to appear in Proceedings of the CKM2006 workshop, Nagoya, Japan.
\bibitem{ref:polci}       F. Polci, M.-H. Schune and A. Stocchi [arXiv:hep-ph/0605129].
\bibitem{ref:kmatrix}   V.V. Anisovich and A.V. Sarantsev, Eur. Phys. Jour. {\bf A16}, 229 (2003).
\end{thebibliography}
\end{document}